# SAveRUNNER: a network-based algorithm for drug repurposing and its application to COVID-19

Giulia Fiscon[1†], Federica Conte[1†], Lorenzo Farina[2], Paola Paci[2*]

1. *Institute for Systems Analysis and Computer Science "Antonio Ruberti", National Research Council, Rome, Italy*

2. *Department of Computer, Control and Management Engineering, Sapienza University of Rome, Rome, Italy*

** Corresponding author*

†*Equal contributors*

## Abstract

The novelty of new human coronavirus COVID-19/SARS-CoV-2 and the lack of effective drugs and vaccines gave rise to a wide variety of strategies employed to fight this worldwide pandemic. Many of these strategies rely on the repositioning of existing drugs that could shorten the time and reduce the cost compared to de novo drug discovery.

In this study, we presented a new network-based algorithm for drug repositioning, called SAveRUNNER (Searching off-lAbel dRUg aNd NEtwoRk), which predicts drug–disease associations by quantifying the interplay between the drug targets and the disease-specific proteins in the human interactome via a novel network-based similarity measure that prioritizes associations between drugs and diseases locating in the same network neighborhoods. Specifically, we applied SAveRUNNER on a panel of 14 selected diseases with a consolidated knowledge about their disease-causing genes and that have been found to be related to COVID-19 for genetic similarity (i.e., SARS), comorbidity (e.g., cardiovascular diseases), or for their association to drugs tentatively repurposed to treat COVID-19 (e.g., malaria, HIV, rheumatoid arthritis). Focusing specifically on SARS subnetwork, we identified 282 repurposable drugs, including some the most rumored off-label drugs for COVID-19 treatments (e.g., *chloroquine*, *hydroxychloroquine*, *tocilizumab*, *heparin*), as well as a new combination therapy of 5 drugs (*hydroxychloroquine, chloroquine, lopinavir, ritonavir*, *remdesivir*), actually used in clinical practice.

Furthermore, to maximize the efficiency of putative downstream validation experiments, we prioritized 24 potential anti-SARS-CoV repurposable drugs based on their network-based similarity values. These top-ranked drugs include ACE-inhibitors, monoclonal antibodies (e.g., anti-IFNγ, anti-TNFα, anti-IL12, anti-IL1β, anti-IL6), and thrombin inhibitors.

Finally, our findings were *in-silico* validated by performing a gene set enrichment analysis, which confirmed that most of the network-predicted repurposable drugs may have a potential treatment effect against human coronavirus infections.

## Introduction

The novel coronavirus disease 2019 (COVID-19) is caused by an enveloped positive-strand RNA virus, named SARS-CoV-2, which affects the respiratory system and whose genome has been likened to the previously identified SARS-CoV strain responsible for SARS outbreak in 2003 [1]. Unfortunately, the world population is completely immune-naïve and therefore vulnerable against this new coronavirus, which has rapidly spread becoming a global pandemic with high morbidity and mortality [2–4]. While COVID-19 lockdowns are easing across the world, main concerns come from its arrival in Africa where the countries are unprepared to counteract the storm's outburst of the new coronavirus and 1.2 billion people are at tremendous risk [5]. For example, Kenya has only 200 intensive care beds for its entire population of 50 million, whereas United States has 34 beds for every 100,000 people. Countries, like Mali, have only a few ventilators for millions of people, health facilities are overcrowded and understaffed. Yet, in many countries throughout Africa, people live together in close quarters, often without access to clean running water that makes social distancing and frequent handwashing, the only possible prevention strategy, all but impossible [5]. Given that, the COVID-19 pandemic demands the rapid identification of repurposable drug candidates to fight the disease progression in the short term and to prevent it from happening in the future.

Drug repurposing is a recent drug development strategy used to identify novel uses for drugs approved by the US Food and Drug Administration (FDA) outside the scope of their original medical indication [6]. It aims at establishing whether an 'old drug' can be reused for new therapeutic purposes representing a faster and cheaper alternative to *de novo* drug discovery process, which generally takes 12-15 years and 2-3 billion dollars (from production to approval, passing through the various phases of preclinical and clinical trials) [6]. Thus far, several therapeutic agents have been evaluated for the treatment of COVID-19, but none have yet been shown to be efficacious [7,8]. Currently, the most promising therapeutic candidate, made available under an emergency-use authorization by the FDA, is *remdesivir*. It is an inhibitor of the viral RNA-dependent RNA Polymerase with proven ability to inhibit SARS-CoV-2 in vitro [9]. In fact, according to recent preliminary results of randomized clinical trials, *remdesivir* has been shown to be superior to placebo in shortening the time to recovery in adults hospitalized with COVID-19 and evidence of lower respiratory tract infection [10]. However, despite the use of *remdesivir*, the mortality remains high, indicating that treatment with an antiviral drug alone is not likely to be sufficient. To continue to improve patient outcomes in COVID-19, combinations of antiviral agents or antiviral agents in combination with other therapeutic approaches should be evaluated by future strategies.

Very promising insights comes from the new emerging field of *network medicine* [11,12], which applies tools and concepts from network theory to elucidate the relation between perturbations on the molecular level and phenotypic disease manifestations. The basic premise of this exercise is that the human interactome (i.e., the cellular network of all physical molecular interactions) can be interpreted as a map and diseases as local perturbations [13]. Yet, the molecular determinants of a given disease (*disease genes*) are not to be randomly scattered, but co-localize and agglomerate in specific regions (*disease modules*) of the interactome and perturbations in these disease modules may contribute to the pathobiological phenotype [12]. From a network medicine perspective, also the action of drugs can be interpreted as a local perturbation of the interactome and thus, for a drug to be on-target effective against a specific disease or to cause off-target adverse effects, its target proteins should be within or in the immediate vicinity of the corresponding disease module. Network-based approaches marrying this philosophy can aid in identifying the specific interactome neighborhood that is perturbed in a certain disease and/or for the effect of a certain drug and guide the search for therapeutic targets, identify comorbidities, as well as rapidly detect drug repurposing candidates [14–18].

Here, we presented SAveRUNNER (Searching off-lAbel dRUg aNd NEtwoRk), a new network-medicine-based algorithm for drug repurposing. It constructs a bipartite drug-disease network by quantifying the interplay between the drug targets and the disease-specific proteins in the human interactome via a novel network-based similarity measure that prioritizes associations between drugs and diseases locating in the same network neighborhoods. SAveRUNNER yielded a high accuracy in the identification of well-known drug indications, thus revealing itself as a powerful tool to rapidly highlight potential novel medical indications for various drugs, which are already approved and used in clinical practice, against the new human coronavirus (2019-nCoV/SARS-CoV-2).

## Results

### *Identification of predicted drug-disease associations*

SAveRUNNER algorithm requires in input a list of drug targets and a list of disease genes to evaluate the extent to which a given drug can repositioned to treat a given disease.

In the present study, disease-associated genes were downloaded from Phenopedia [19], which provides a disease-centered view of genetic association studies; whereas drug-target associations were obtained from DrugBank [20], which collects huge amount of drug-related data, recently enabling the discovery and repurposing of a relevant number of existing drugs to treat rare and newly identified diseases [6,14] (Figure 1a).

We tested the performance of SAveRUNNER on a panel of 14 diseases that have been found to be related to COVID-19 for genetic similarity, comorbidity, or for their association to drugs with the potential to shorten the recovery time for seriously ill COVID-19 patients. Thus, we included Severe Acute Respiratory Syndrome (SARS) since it is caused by the coronavirus with the highest nucleotide sequence identity with SARS-CoV-2 [16,21]. Unlike COVID-19, for which there is still a partial knowledge of its associated disease genes, SARS has been widely studied and a reliable list of the its molecular determinants is available in the most common databases of human genetic associations [19]. Yet, we comprehended cardiovascular diseases, diabetes and hypertension, whose comorbidity in COVID-19 patients is well documented [22,23]. In addition, we included other viral infections (i.e., malaria, HIV and Ebola) and immune disorders (i.e., rheumatoid arthritis), since drugs that have been authorized for their treatment (i.e., *chloroquine/hydroxychloroquine, lopinavir/ritonavir*, *remdesivir*, and *tocilizumab*, respectively) are being investigated worldwide for their potential to treat coronavirus disease (COVID-19) [7,9,24–30].

For what concerns drug-target associations, we assembled target information for a total of 1875 FDA-approved drugs. In addition, we considered in our input list of drug targets also the combination of *remdesivir* with other four antiviral agents, i.e. *hydroxycloriquine*, *chloriquine*, *lopinavir*, and *ritonavir* (referred as "5-cocktail"), as well as the combination of two of them, i.e. *lopinavir* and *ritonavir* (referred as "kaletra"), whose antiviral action against coronavirus infections has been demonstrated both *in vitro* and *in vivo* studies [31,32]. In fact, despite significant progress in the COVID-19 management pointing to *remdesivir* as the most promising therapeutic candidate, the mortality remains high, indicating that treatment with an antiviral drug alone is not likely to be sufficient and combinations of antiviral agents should be evaluated by future strategies to improve patient outcomes in COVID-19 [10].

The complete lists of the analyzed diseases and drugs are provided in Supplementary Table 1.

The rationale behind SAveRUNNER algorithm lies on the hypothesis that, for a drug to be effective against a specific disease, its associated targets (drug module) and the disease-specific associated genes (disease module) should be nearby in the human interactome [14] (Figure 1b). To quantify the vicinity between drug and disease modules, SAveRUNNER implements a novel network similarity measure:

$$f(p) = \frac{1}{1 + e^{-c\left[\frac{(1+QC)(m-p)}{m} - d\right]}}$$

where $p$ is the network proximity measure defined in [14]:

$$p(T,S) = \frac{1}{\|T\|}\sum_{t\epsilon T} \min_{s\epsilon S} d(t,s)$$

that represents the average shortest path length between drug targets $t$ in the drug module $T$ and the nearest disease genes $s$ in the disease module $S$; $QC$ is the quality cluster score; $m$ is $max(p)$; $c$ and $d$ are the steepness and the midpoint of $f(p)$, respectively (see Materials and Methods). To assess the statistical significance of this new defined network-based similarity measure, SAveRUNNER applies a degree-preserving randomization procedure, expecting a p-value $\leq 0.05$ for proximal drug and disease modules.

The novelty of our approach resides in having implemented a procedure to prioritize the predicted off-label drug indications for a given disease (see Materials and Methods section). This prioritization procedure exploits a clustering analysis to reward associations between drugs and diseases belonging to the same network cluster, based on the assumption that if a drug and a disease group together is more likely that the drug can be effectively repurposed for that disease. In this sense, we say that drugs and disease that are members of the same group (cluster), are more similar to each other than to members of other groups (clusters).

SAveRUNNER algorithm releases as output a weighted bipartite drug-disease network, in which one set of nodes corresponds to drugs and the other one corresponds to diseases. A link between a drug and a disease occurs if the corresponding drug targets and disease genes are nearby in the interactome more than expected by chance (p-value $\leq$ 0.05) and the weight of their interaction corresponds to the new defined network-based similarity measure. In this study, the final drug-disease network was composed of a total of 1682 nodes (i.e., 14 diseases associated to 1668 drugs) and 7177 edges.

This drug-disease network was naturally rendered as a matrix reporting the 14 diseases on the rows, and the 1668 drugs on the columns (Figure 2), labeled with their original medical indication according to the Therapeutic Target Database (TTD). Each matrix cell was colored according to the corresponding similarity value of a given drug-disease pair: shades of yellow denote drug-associated targets more proximal (high similarity) to the disease-associated genes in the human interactome, whereas shades of blue denote drug targets more distal (low similarity) to the disease genes (Figure 2).

For elucidating drugs/diseases relatedness in terms of network similarity, we computed a hierarchical biclustering on the drug-disease similarity matrix. This analysis pointed out two main disease clusters: one including SARS, RDS, multiple sclerosis, rheumatoid arthritis, malaria, and viral infection diseases (magenta box in Figure 2); the other one including cardiovascular diseases and their risk factors, i.e., diabetes mellitus and hypertension (green box in Figure 2).

As proof of validity of SAveRUNNER, among the predicted drug-disease associations, we found several already known associations. For example, *tocilizumab*, *interferon-beta 1a/b*, *chloroquine/ hydroxychloroquine*, *lopinavir*/*ritonavir*, resulted to be significantly associated (p-value $\leq$ 0.05) with the diseases for which they were approved (i.e. rheumatoid arthritis, multiple sclerosis, malaria, and HIV infection, respectively).

### *SARS-CoV-host interactome*

In total, we found 41 host proteins associated with SARS-CoV and their specific subnetwork within the human interactome is shown in Figure 3a.

By mapping the known drug–target network into the SARS-CoV-host interactome, we revealed that 21 out of 41 (51%, light blue nodes in Figure 3a) SARS-CoV-associated disease genes are druggable cellular targets, i.e. can be directly targeted by at least one FDA-approved drug. For example, F2, TNF, FCGR2A, ACE, are the most targetable proteins. The KEGG pathway enrichment analysis revealed that this 21 targetable SARS-CoV-associated disease genes are involved in multiple significant (adjusted p-value $\leq$ 0.05) viral infection-related pathways, including TNF signaling pathway, Toll-like receptor signaling pathway, Cytokine-cytokine receptor interaction, IL-17 signaling pathway, Epstein-Barr virus infection, Inflammatory bowel disease, and Influenza (Figure 3b), mirroring the enrichment results found for the total 41 SARS-CoV-associated disease genes (Figure 3c).

The high druggability of SARS-CoV-host interactome and the relevance of the pathways in which the 21 druggable SARS-CoV-associated disease genes are involved strongly supports a drug repurposing strategy for potential treatment of COVID-19, by specifically targeting cellular proteins associated with SARS.

The remaining 20 non-targetable SARS-CoV-associated disease genes include proteins (e.g., CD14, IFNA1, IFNB1, IL4, IL10, CCL5) having key roles in the immune-inflammatory response to viral infection as well. This strongly supports a network-based view of drug action, according to which most disease phenotypes are difficult to reverse through the use of a single “magic bullet”, that is, an intervention that affects a single node in the network and encourages to inspect all the molecules that are situated in the neighborhood of the targetable proteins, leading to also detect possible side-effects [12].

### *SARS-CoV drug-disease network*

In total, SAveRUNNER identified 282 repurposable drugs that were significantly associated (p-value $\leq$ 0.05) with the SARS-CoV infection (Supplementary Table 2). Figure 4a illustrates a sketch of SARS drug-disease network in which SARS is connected to the other analyzed diseases via the high-confidence predicted drugs (with their original medical indications) that could be repurposed for potential SARS fighting. This network is composed of two independent sets of nodes: one set corresponds to 13 diseases (i.e., SARS and the other 12 diseases sharing at least one drug with SARS) represented as red circles and whose size scales with their degree; the other one corresponds to the 66 FDA-approved non-SARS drugs for which known therapeutic indications were available from TTD or to the drugs combination proposed as new potential medical indication, colored accordingly. Two nodes of the two independent sets were connected whether the network similarity of the drug targets and disease genes in the human interactome was statistically significant (p-value $\leq$ 0.05).

Looking at Figure 4a, it clearly emerges that there exist several candidate repurposable drugs for treatment of SARS. Among them, we pointed out the 5-cocktail (adjusted similarity = 0.89, p-value = 0.036) that appeared as a promising candidate for targeting SARS showing a quite high network similarity.

In total, among the candidate repurposable drugs for SARS with an already known medical indication, we found 16 drugs indicated for hypertension treatment, including *enalapril* which is also the top potential anti-SARS-CoV repurposable drug (adjusted similarity = 0.99, p-value = $1.21\ 10^{-32}$); 14 drugs for cardiovascular diseases, including *heparin* (adjusted similarity = 0.19, p-value = 0.0046); 11 drugs for rheumatoid arthritis, including *tocilizumab* (adjusted similarity = 0.99, p-value = $6.5\ 10^{-4}$); 7 drugs for respiratory diseases; 6 drugs for diabetes mellitus; 4 drugs for multiple sclerosis, including *interferon beta-1a* (adjusted similarity = 0.99, p-value = $8.22\ 10^{-7}$) and *interferon beta-1b* (adjusted similarity = 0.99, p-value = $2.18\ 10^{-16}$); 4 drugs for HIV infection; 3 drugs for malaria, including *chloroquine* (adjusted similarity = 0.35, p-value = 0.01) and *hydroxycloriquine* (adjusted similarity = 0.51, p-value = $1.85\ 10^{-6}$); and 2 drugs for influenza (Figure 4b).

By analyzing the distribution of all the known medical indications associated to the drugs shared between SARS infection and each analyzed disease, hypertension- and cardiac-associated drugs appeared as highly frequent drugs, followed by rheumatoid arthritis- and respiratory disease-associated ones (Figure 4c). This findings appear in accordance with the fact that patients with SARS-CoV-2 infection also showed potential hypertension and cardiac injuries, including arrhythmia and myocardial dysfunction [22,23,33,34]. This could be owed to the disruption of the subtle balance between angiotensin-converting enzyme 1 (ACE1) and angiotensin-converting enzyme 2 (ACE2), identified as functional receptor of both SARS-CoV/SARS-CoV-2 for host cell entry [35] and normally devoted to present various cardiovascular protective effects [33].

### *GSEA validation*

In order to validate the 282 anti-SARS repurposable drugs predicted by SAveRUNNER, we performed a gene set enrichment analysis (GSEA) as in [16], by using transcriptome data of SARS-CoV infected cells as gene signature for SARS. The gene expression data of drug-treated human cell lines from the Connectivity Map (CMAP) database were exploited to obtain drug signatures and thus to calculate a GSEA score for each drug as an indication of *in-silico* validation. In particular, for each SARS dataset, we selected drugs with a score > 0 in order to focus on those drugs able to have a potential treatment effect on genes that are hallmark for that phenotype. The assigned GSEA score, ranging from 0 to 2, corresponded to the number of datasets satisfying this criterion for a specific drug.

The GSEA analysis validated a total of 61 out the 282 candidate drugs to be repositioned against SARS infection, including 26 with a GSEA score of 2 and 35 with a GSEA score equal to 1 (Supplementary Table 2).

### *SARS-CoV-2 host interactome*

In this study, we focused on SARS disease, which is caused by the coronavirus having the higher genetic similarity with SARS-CoV-2 [16,21] and for which the specific disease genes are known. Conversely, given the novelty of the new coronavirus COVID-19 in the spectrum of human disease, the knowledge of the COVID-19-associated genes is far from completeness. However, a study has been recently published, where the authors identified 332 human proteins interacting with 26 SARS-CoV-2 proteins by using affinity purification mass spectrometry [36]. Although the authors verified that these proteins were preferentially highly expressed in lung tissue (typical environment where the virus causes a major damage), this study has been carried out on human HEK293T kidney cells that does not represent the primary physiological site of infection.

For a more generalizable understanding of human coronaviruses infection, we completed our analysis by applying SAveRUNNER on these the 332 human proteins preliminarily associated to SARS-CoV-2. The obtained results are encouraging and appears in accordance with the analysis carried out on SARS-CoV. Indeed, by performing a hierarchical clustering on the network of predicted drug-disease associations, we found COVID-19 in the same cluster of SARS, RDS, multiple sclerosis, rheumatoid arthritis, malaria, and viral infection diseases (Figure 5a). This is in accordance with recent studies that are attempting to repurpose, for COVID-19 treatment, drugs approved to treat other viral infections such as influenza, malaria, HIV, and Ebola or immune-related disorders, such as rheumatoid arthritis and multiple sclerosis [9,24–26,28,29,37].

The list of network-predicted drugs potentially able to treat SARS-CoV-2 infection contains a total of 98 drugs, including 54 (i.e., 55%) COVID-19 specific and 44 (i.e., 45%) in common with the 282 candidate repurposable drugs found for SARS-CoV (Supplementary Table 2 and Figure 5b). The high number of repurposing candidates shared between SARS-CoV-2 an SARS-CoV is also reflected in the result of the greedy clustering algorithm implemented by SAveRUNNER, which places COVID-19 and SARS together in the cluster with the highest quality cluster score.

We then investigated whether the network-predicted repurposable drugs could counteract the gene expression perturbations caused by SARS-CoV-2, i.e. if they could up-regulate genes down-regulated by the infection or *viceversa*. To verify that, we performed the GSEA analysis by using, as COVID-19 gene signature, the 120 differentially expressed genes in SARS-CoV-2-infected A549 cells reported in [38] and, as drug signatures, the gene expression data of drug-treated human cell lines from the CMAP database [39]. Interestingly, this analysis revealed that 24 out of 98 candidate drugs to be repositioned for COVID-19 treatment achieved the highest GSEA score. In particular, 14 of them, encompassing *lopinavir*, were included in the 54 specific-SARS-CoV-2 predicted drugs (Supplementary Table 2 and Figure 5b).

In a recent study [18], the authors exploited the new knowledge of SARS-CoV-2 host interactome [36] and integrated several network-based drug repurposing strategies to prioritize 81 promising repurposing candidates against COVID-19. The overlap between these 81 repurposable drugs and the 98 ones predicted by SAveRUNNER is of 5 drugs, i.e. *isoniazid, lopinavir, romidepsin, sulfinpyrazone, tadalafil.*

### *Prediction of disease comorbidity*

In order to predict potential comorbidity patterns among all the diseases analyzed in this study, for each disease module corresponding to the 15 analyzed disorders, we computed the non-Euclidean separation distance, which measures the modules' overlap [40]:

$$s(A,B) = p_{AB} - \frac{p_{AA} + p_{BB}}{2}$$

where $p(A,B)$ is the network proximity defined as:

$$p(A,B) = \frac{1}{|A| + |B|}\left[\sum_{a\epsilon A} \min_{b\epsilon B} d(a,\, b) + \sum_{b\epsilon B} \min_{a\epsilon A} d(b,\, a)\right]$$

and $d(a,b)$ is the shortest distance between disease gene $a$ of module $A$ and disease gene $b$ of module $B$. A positive value for the separation measure indicates that two disease modules are topologically well-separated in the human interactome, whereas a negative value for the separation measure indicates that two disease modules are located in the same network neighborhood and thus overlap.

We found that SARS-CoV disease module directly overlapped with RDS and Influenza disease modules ($s < 0$) and appeared to be closest to Malaria (smallest $s$) followed by cardiomyopathies. (Figure 6a). On the other hand, by using SARS-CoV-2 disease genes, we found that the COVID-19 disease module did not directly overlap with any disease module here analyzed (Figure 6b). However, among the closest diseases (smallest $s$), we found cardiovascular diseases (i.e., cardiomyopathies, arrhythmia, heart arrest), whose comorbidity in COVID-19 patients has been already discussed [22,23].

## *Comparison with other methods*

In the last few years, several computational network-based methods have been proposed to predict direct drug–disease associations for drug repositioning [41–46]. Among them, the MBiRW algorithm adopts adopted an effective mechanism to measure similarity for drugs and diseases and applied a Bi-Random walk (BiRW) algorithm to predict potential new indications for existing drugs [46]. This methodology has been shown to outperform other well-known network-based prediction methods [15,47,48] in correctly predicting true drug–disease associations. These captivating results prompted us to implement a BiRW-based algorithm (see Materials and Methods) against which we compared the performance of SAveRUNNER.

The effectiveness of the drug–disease predictions provided by BiRW and SAveRUNNER were evaluated and compared in terms of Receiver Operating Characteristic (ROC) probability curves with their corresponding Area Under the Curve (AUC) (see Materials and Methods). In particular, we found that SAveRUNNER yielded over 70% accuracy (AUC = 0.73) for identifying well-known drug-disease relationships and overcame the one obtained by the BiRW-based algorithm (AUC = 0.59). In other words, there is 73% chance that SAveRUNNER algorithm will be able to distinguish between positive class (known drug-disease associations) and negative class (unknown drug-disease associations) against the 59% of the BiRW-based algorithm.

## Discussion

### *Ongoing clinical trials for the management of COVID-19*

Thus far, no proven effective therapies for the novel coronavirus disease COVID-19 exist and the majority of available data are based on expert opinions and anecdotal experiences [49]. Old and new agents have been proposed and explored for treatment of COVID-19 [50], but clinical trials are still underway. Currently, 291 active trials specific to COVID-19 are present on Clinical-Trials.gov. We will next discuss some of the most popular ongoing clinical trials for the management of the COVID-19 pandemic [7,9,27,51–56].

*Remdesevir* is a novel nucleotide analogue currently under evaluation in clinical trials for Ebola infection and it has shown an excellent activity against early coronavirus infections (SARS, MERS) both *in vitro* and in animal models. Recently, Grein J et al. [56] provided *remdesivir* on a compassionate-use basis to patients hospitalized with COVID-19. In this cohort of patients, clinical improvement was observed in 36 of 53 patients (68%). Although *remdesivir* has not yet been approved by US Food and Drug Administration, a preliminary report published on May 22, shows that it was superior to placebo in shortening the time to recovery in adults with COVID-19 and evidence of lower respiratory tract infection [10].

*Chloroquine* is an old and widely used anti-malarial drug and it is also efficacious as an anti-inflammatory agent for rheumatologic disease. Earlier studies have demonstrated a potential antiviral effect of chloroquine that may depend by several mechanisms such as the change of cell membrane pH, which is necessary for viral fusion and the interference with glycosylation of viral proteins. A recent study has demonstrated *in vitro* efficacy of *chloroquine* and *remdesivir* in inhibiting replication of SARS-CoV-2 [9]. Moreover, emerging reports from China suggest that *chloroquine* has shown a superiority in reducing both the severity and the duration of clinical disease without significant adverse events in almost one hundred patients [26,51,52]. In light of this results, an expert consensus group in China has recommended *chloroquine* for COVID-19 treatment [26].

*Hydroxychloroquine* (brand name plaquenil) is an analogue of *chloroquine*, which was demonstrated to be much less toxic than *chloroquine* in animals and with similar *in vitro* efficacy on SARS-CoV-2 [25,27]

*Lopinavir*/*ritonavir* (brand name *kaletra*) is a well-known protease inhibitor, which has been widely used for many years for the treatment of HIV infection. Compared to *remdesivir*, *lopinavir*/*ritonavir* has the advantage that it is widely available and has an established toxicity and drug-drug interactions profile. Its antiviral action against coronavirus infections has been previously demonstrated both *in vitro* and *in vivo* (animal and human data) studies conducted on SARS [31,32]. Recently Cao et al. [7] conducted a randomized, controlled, open-label trial involving hospitalized adult patients with confirmed SARS-CoV-2 infection. A total of 199 patients with laboratory-confirmed SARS-CoV-2 infection underwent randomization; 99 were assigned to the *lopinavir*/*ritonavir* group, and 100 to the standard-care group. Unfortunately, the trial results were disappointing and no benefit was observed with *lopinavir*/*ritonavir* treatment in hospitalized adult patients with severe COVID-19. Nevertheless, *lopinavir*/*ritonavir* is currently under investigation within other randomized clinical trials and the results of such trials will provide convincing positive or negative findings on this therapy.

In absence of proven antivirals for COVID-19, adjunctive therapies of support represent the cornerstone of care. In particular: anticytokine drugs, corticosteroids, low molecular weight heparin (LMWH), and immunoglobulin therapy. Monoclonal antibodies directed against key inflammatory cytokines or other aspects of the innate immune response represent another potential class of adjunctive therapies for COVID-19.

One of the most promising monoclonal antibodies under investigation for the management of COVID-19 is *tocilizumab* (TCZ). Specifically, TCZ is an anti-human IL-6 receptor monoclonal antibody that inhibits signal transduction by binding sIL- 6R and mIL-6R. Currently, TCZ is licensed for the treatment of adult patients with moderately to severely active rheumatoid arthritis, but several studies in China showed a possible correlation of massive inflammation and severe lung damage on the rapid evolution of fatal pneumonia. Indeed, in COVID-19 patients, significant differences in IL-6 plasmatic levels were observed at different stage of disease with a higher expression in severe cases [37]. Despite the lack of clinical trials on TCZ efficacy and safety for COVID-19 treatment, “off-label” TCZ has been used as potential treatment strategy in severe and critical COVID-19 patients. Currently, in Italy a multicenter study on the efficacy and tolerability of tocilizumab in the treatment of patients with COVID-19 pneumonia is ongoing.

Another remarkable candidate for the treatment of COVID-19 is LMWH, with the aim to improve the coagulation dysfunction of COVID-19. Really, during the course of SARS-CoV-2 infection, an increased incidence of acute pulmonary embolism episodes was detected. These were often COVID-19 patients with risk factors for embolism, increased D-dimer and deterioration of general conditions and respiratory failure. Recently, Tong N. et al. [53] showed that anticoagulant therapy mainly with LMWH appears to be associated with better prognosis in severe COVID-19 patients meeting criteria of the sepsis-induced coagulopathy score or with markedly elevated D-dimer. However, its efficacy remains to be validated in large clinical trials. Interestingly, LMWH seems exert anti-inflammatory effects by reducing IL-6 and increasing lymphocyte %, so that the potential of LMWH could be to mitigate cytokine storm in severe COVID-19 patients.

To complete the clinical landscape of potential adjunctive therapy for COVID-19, we mention also the use of convalescent plasma or hyperimmune immunoglobulins obtained from recovered patients [57]. The rationale for this treatment is that antibodies from recovered patients may help the immune clearance of SARS-CoV-2.

In conclusion, several drugs demonstrate *in vitro* activity against SARS-CoV-2. Of these, several repurposed agents used to treat a variety of other diseases (malaria, HIV, rheumatoid arthritis) have been proposed as possible cure options for COVID-19. *Lopinavir/ritonavir* and *chloroquine* or *hydroxychloroquine* are the treatments with the most clinical evidence, either positive or negative, in the treatment of COVID-19. Currently, randomized clinical trials have not proven that any of these drugs are undoubtedly effective.

### *In-silico drug predictions for the management of COVID-19*

In this study, we developed a novel network-based algorithm for drug repurposing, called SAveRUNNER, with the aim to offer a promising framework to efficiently screen potential novel indications for currently marketed drugs against COVID-19.

Our findings, in accordance with several recent works [14,16,58], suggested that the discovery of efficacious repurposable drugs (or drug combinations) could benefit from the exploration of the relationship between drug targets and the disease genes in the human interactome. The novelty of SAveRUNNER relies on the definition of a new network-based similarity measure, which quantifies the vicinity between drug and disease modules and considers the drug-disease network modular structure to reward predicted associations between drugs and diseases that are located in the same network neighborhoods.

Focusing on SARS disease, which is caused by the coronavirus having the higher genetic similarity with SARS-CoV-2 [16,21] and for which the specific disease genes are known, we identified 282 candidate repurposable drugs. Among them, we recovered some of the most rumored off-label drugs, like *chloroquine, hydroxycloriquine, tocilizumab,* and *heparin*. While *lopinavir/ritonavir* and *remdesivir* were not found to be significantly associated with SARS, their combination together with *chloroquine* and *hydroxycloriquine* (here referred as 5-cocktail) was predicted as a promising anti-SARS-CoV repurposable drug.

Although all the 282 repurposable drugs predicted by SAveRUNNER warrant to be explored, a prioritization of these drugs according to the decreasing value of their network similarity value with SARS may offer the possibility to maximize the efficiency of subsequent experimental screening and clinical trial validation. This does not mean that drugs located at a lower-ranking have no potential efficacy or must be *a priori* excluded from further exploration.

Thus, we found that the 95$^{th}$ percentile of the similarity values distribution includes ACE-inhibitors (i.e., *enalapril, trandolapril, fosinopril, benazepril, cilazapril, zofenopril, spirapril, rescinnamine, quinapril*), thrombin inhibitors (i.e., *thrombomodulin alfa, bivalirudin, dabigatran etexilate, argatroban, ximelagatran*), and several monoclonal antibodies like: anti-TNFα (i.e., *adalimumab, golimumab, infliximab, certolizumab pegol*), anti-IFNγ (i.e., *emapalumab*), anti-IL1β (i.e., *canakinumab*), and anti-IL6 (i.e., *siltuximab*).

These findings are in accordance with the current therapeutic avenues tentatively proposed for fighting COVID-19, even if, for some of them, the effectiveness in treating COVID-19 remains controversial. Indeed, several recent studies hypothesized that COVID-19 patients receiving ACE-inhibitors may be subject to poorer outcomes [59,60], whereas other investigators argued that the usage of ACE-inhibitors could be beneficial in COVID-19 infection [61]. On the contrary, the potential efficacy of the other top-ranked predicted off-label drugs seems to be less uncertain. Indeed, several evidences suggested that COVID-19 may predispose patients to arterial and venous thrombotic disease and then common antithrombotic medications, including the already mentioned *heparin* or direct thrombin inhibitors such as *dabigatran,* have been proposed as potential adjunctive therapy against COVID-19 [62]. Yet, monoclonal antibodies targeting key inflammatory cytokines or other aspects of the innate immune response are increasingly recognized as another promising class of anti-COVID-19 drugs. In particular, the class of anti-TNFα antibodies, mostly used for the treatment of inflammatory rheumatic diseases, could be able to challenge COVID-19 by two main actions: the classical TNFα inhibition and a down-regulation of ACE2 expression resulting in decreased binding sites for SARS-CoV-2 [63,64]. Thus, anti-TNFα antibodies, and *adalimumab* in particular thanks to its excellent safety profile [65], could inhibit the basic mechanisms of COVID-19, and could be potentially useful in managing/preventing COVID-19-driven pneumonia[66]. As proof of this, a study evaluating *adalimumab* injection in COVID-19 patients has recently been registered in order to assess the role of this antibody in treating COVID-19 patients with severe pneumonia [67].

### *In-silico validation of SAveRUNNER predictions*

As indication of *in-silico* validation of the 282 candidate repurposable drugs predicted by SAveRUNNER, we performed a GSEA analysis that allowed to investigate whether these repositioning candidates have potential treatment effect against SARS-CoV infection. The GSEA analysis validated a total of 61 out the 282 anti-SARS repurposable drugs. Among the drugs achieving highest GSEA scores, we found some of the ACE-inhibitors above-mentioned (i.e., *enalapril, benazepril, quinapril, fosinopril, rescinnamine*) as well as ACE-inhibitors with multiple targets (i.e., *perindopril, captopril, moexipril, ramipril*), and *chloroquine*.

In addition, this analysis came up other interesting but less obvious drugs falling far outside antiviral use such as *ruxolitinib, lovastatin*, and a group of H1-antihistamines (i.e., *loratadine, fexofenadine, levocetirizine, desloratadine, clemastine, ketotifen, diphenhydramine, cetirizine*).

*Ruxolitinib* is a potent and selective Janus-Associated Kinase (JAK) inhibitor approved for treatment of myelofibrosis, but thanks to its powerful anti-inflammatory effect, it could be likely to be effective against the consequences of the elevated levels of cytokines typically observed in patients with COVID-19 [68,69].

*Lovastatin* is widely prescribed to reduce the levels cholesterol in the blood and prevent cardiovascular diseases. Experimental evidence suggested that statin-induced reduction of cholesterol in the plasma membrane results in lower viral titers and failure to internalize the virus [70].

H1-antihistamines are mostly used to treat allergic reactions, but they potentially could have beneficial anti-inflammatory effects on immune dysregulation during COVID 19 infection. Indeed, histamine is a biologically active substance that potentiates the inflammatory and immune responses of the body, regulates physiological function in the gut, and acts as a neurotransmitter.

Notably, among the repurposable drugs against SARS-CoV and SARS-CoV-2 both predicted by SAveRUNNER and validated by GSEA analysis, we found 10 common off-label drugs, encompassing ACE inhibitors (i.e., *enalapril*, and *quinapril*) and antihistamines (i.e., *cetirizine* and *fexofenadine*). Moreover, among repurposable drugs validated by the GSEA analysis as potential treatment effect against SARS-CoV-2 infection, we found the protease inhibitor *lopinavir*.

### *Drugs' likely mechanism-of-action in SARS*

Moving beyond the "one drug, one target" vision prompted us to explore all the molecules that are situated in the interactome neighborhood of the drugs targetable proteins, which may be altered by the drug activity as well as may cause side-effects (Figure 7).

We interestingly observe that all drug subnetworks include genes involved in the immune-inflammatory response and share individual-specific genetic factors (i.e., HLA-A, HLA-B, HLA-C), as well as some crucial pro-inflammatory cytokines (i.e., IFN-γ, IL-1β, TNF). The HLA genes encodes for the proteins of the major histocompatibility complex of class I (MHC-I), that is directly involved in the antigen presentation process. The cytokines IFN-γ, IL-1β, and TNF have a critical role in promoting inflammation [71,72] and several studies demonstrated as their suppression may lead to therapeutic effects in many inflammatory diseases, including viral infections [66,73–75]. An overwhelming production of these pro-inflammatory cytokines indeed contributes to a hyper-inflammatory condition, denoted as *cytokine storm*, which destroys the normal regulation of the immune response and may induce pathological inflammatory disorders [75–77].

Some drugs are characterized by highly complex interaction networks, whereas others, like the case of *remdesivir*, seem to have a poorly connected substructure. However, it is worth noting that the observed network complexity does not scale with the potential drug effectiveness. The lack of interactions may be likely ascribable both to the current incompleteness of the human interactome and to the shortage of drug-target information. As in the case of *remdesivir*, which being designed to target specifically virus proteins, little is known about its human targetable proteins, leading to a less wired subgraph.

This is not the case of the *heparin* subnetwork (Figure 7f), where the highly complexity in its interaction network perfectly matches its versatile role of anticoagulant, anti-inflammatory, and antiviral drug turning out in a strong multifactorial impact in the new human coronavirus [55]. In fact, the *heparin* subnetwork includes: disease genes associated to the complement and coagulation cascades pathway (MASP2, MBL2), confirming its primary anticoagulant function; pro-inflammatory cytokines (IL12B and its receptor IL12RB1) and chemokines (CXCL5, CXCL9, CXCL10), supporting its anti-inflammatory function; all the disease genes of the antiviral drugs subnetworks (Figure 7a-d), linking to its antiviral action.

Yet, the networks of JAK inhibitors such as *ruxolitinib* (Figure 7i) and of the H1-antihistamines (Figure 7l) strongly confirm their anti-inflammatory effect, showing common and specific key players of the inflammatory response. In particular, *ruxolitinib* subnetwork includes several pro-inflammatory cytokines (IFN-α, IFN-γ, IL-1β, IL-6, IL-12, TNF-α), chemokines (CCL2, CCL5), and other important specific interleukines (IL-10, IL-4); whereas the H1-antihistamines network shows how the drug may control the level of the interleukin IL-6, directly linked to the histamine receptor H1 (HRH1).

## Materials and Methods

### *Human protein–protein interactome*

The human protein–protein interactome was downloaded from the Supplementary Data of [14], where the authors merged their inhouse systematic human protein–protein interactome and 15 commonly used databases with several types of experimental evidences (e.g., binary PPIs from three-dimensional protein structures; literature-curated PPIs identified by affinity purification followed by mass spectrometry, Y2H, and/ or literature-derived low-throughput experiments; signaling networks from literature-derived low-throughput experiments; kinase-substrate interactions from literature-derived low-throughput and high-throughput experiments). This updated version of the human interactome is composed of 217,160 protein–protein interactions (edges or links) connecting 15,970 unique proteins (nodes).

### *Disease-gene associations*

Disease-associated genes were downloaded from Phenopedia [19], which provides a disease-centered view of genetic association studies collecting by the online Human Genome Epidemiology (HuGE) encyclopedia [78].

The updated version of Phenopedia (released 27-04-2020) collects gene associations for 3,255 diseases. Among them, we selected a panel of 14 diseases of interest with their associated genes (Supplementary Table 1).

### *Drug-target interactions and drug medical indications*

Drug-target interactions were acquired from DrugBank [20], which is a comprehensive, freely accessible, online database containing information on drugs and drug targets. The updated version of DrugBank (version 5.1.6, released 22-04-2020) contains 13,563 drug entries including 2,627 approved small molecule drugs, 1,373 approved biologics (proteins, peptides, vaccines, and allergenics), 131 nutraceuticals, and over 6,370 experimental drugs. For our analysis, we selected a total of 1,875 FDA-approved drugs with at least one annotated target. The target Uniprot IDs were mapped to Entrez gene IDs by using BioMart – Ensembl tool (https://www.ensembl.org/). Note that, for some drugs of interest for which no targets were found in DrugBank, we integrated drug-target interactions available from Therapeutic Target Database (TTD) [79] and Pharmacogenomics Knowledgebase (PharmGKB) [80] database. In particular, for remdesivir (an anti-virus drug designed to target specifically virus proteins), we extracted its human target information from TTD.

The known drug medical indications were obtained from TTD [79], whose last version was released on 11 Nov 2019.

### *SAveRUNNER algorithm*

SAveRUNNER (Searching off-lAbel dRUg aNd NEtwoRk) is a network-based algorithm for drug repurposing that, taking as input a list of drug targets and disease genes, constructs a drug-disease network with predicted drug-disease associations by performing the steps that will be next discussed (Figure 8).

#### **Step 1: Computation of network proximity**

In order to investigate the extent to which disease and drug modules are close in the human interactome, SAveRUNNER implements the network-based proximity measure defined as [14]:

$$p(T,S) = \frac{1}{\|T\|} \sum_{t \epsilon T} \min_{s \epsilon S} d(t,s)$$

which is the average shortest path length between drug targets $t$ in the drug module $T$ and the nearest disease genes $s$ in the disease module $S$ (Figure 1b). To evaluate the statistical significance of the observed network proximity between the two modules $T$ and $S$, SAveRUNNER builds a reference distance distribution corresponding to the expected distance between two randomly selected groups of proteins with the same size and degree distribution of the original sets of disease proteins and drug targets in the human interactome. This procedure is repeated 1,000 times and the z statistics, together with the corresponding p-value, is computed by using the mean and the standard deviation of the reference distance distribution.

**Step 2: Computation of network similarity**

The network proximity measure is then translated in a similarity measure assuming values in the range [0-1]:

$$similarity = \frac{\max(p) - p}{\max(p)}$$

where null similarity means that the corresponding disease and drug modules are very distal in the human interactome (i.e., $p$ is maximum); whereas maximum similarity means that the corresponding disease and drug modules are very proximal in the human interactome (i.e., $p$ equal to zero).

**Step 3: Selection of p-value threshold**

In order to filter out statistically insignificant drug-disease associations, a significance level for the p-values is set. It means that, given a disease *A* and a drug *b*, if the p-value associated to their distance in the human interactome is smaller of the chosen significance level, the probability that the off-label drug *b* would be effective for this disease *A* is greater than expected by chance. In our analysis, predicted drug-disease associations with a p-value $\leq 0.05$ were selected.

**Step 4: Cluster detection**

Next, SAveRUNNER performs a cluster analysis to detect groups of drugs and diseases in such a way that members in the same group (cluster) are more similar to each other than to those in other groups (clusters). To do that, SAveRUNNER exploits a cluster detection algorithm based on the greedy optimization of the network modularity [81]. Specifically, modularity measures the strength of network division into clusters, i.e.

networks with high modularity have dense connections between the nodes within clusters but sparse connections between nodes in different clusters. The greedy clustering algorithm was adopted in view of its good performance and high speed in detecting community structure from very large networks [81]. The quality of each cluster is evaluated by SAveRUNNER by computing the following quality cluster ($QC$) score:

$$QC = \frac{W_{in}}{W_{in} + W_{out} + P}$$

where $W_{in}$ denotes the total weight of edges within the cluster, $W_{out}$ denotes the total weight of edges connecting this cluster to the rest of network, and $P$ is a penalty term which considers the node density within the cluster.

### Step 5: Adjustment of network similarity

The quality cluster score serves to reward associations between drugs and diseases belonging to the same cluster, based on the assumption that if a drug and a disease group together is more likely that the drug can be effectively repurposed for that disease. In this sense, we say that drug and the disease that are members of the same cluster tend to be "more similar" and this translates into the following adjustment for the similarity:

$$similarity = (1 + QC) \cdot similarity \tag{1}$$

Thus, whether two nodes fall in the same cluster their similarity value increases by a factor proportional to the $QC$ score of the cluster which they belong; otherwise whether two nodes do not fall in the same cluster $QC$ is set to zero and their similarity value does not change.

### Step 6: Normalization of network similarity

To bound the similarity measure defined in Eq. 1 to values that monotonically increase from 0 to 1, SAveRUNNER performed a normalization procedure by applying the following sigmoid function:

$$f(x) = \frac{1}{1 + e^{-c(x-d)}}$$

where $x$ is the adjusted similarity measure (Eq. 1), $d$ is the sigmoid midpoint (i.e., the value at which the function approaches to 0.5), $c$ is the sigmoid steepness. We set $d = \frac{\max(x)}{2}$ and $c = 10$.

At the end of this step, SAveRUNNER offers a list of predicted/prioritized associations between drugs and diseases as a weighted bipartite drug-disease network, in which one set of nodes corresponds to drugs and the other one corresponds to diseases. A link between a drug and a disease occurs if the corresponding drug targets and disease genes are nearby in the interactome more than expected by chance (p-value $\leq 0.05$) and the weight of their interaction corresponds to the adjusted and normalized similarity value.

### *Performance evaluation*

The effectiveness of predicted drug–disease associations provided by the SAveRUNNER algorithm was evaluated in terms of the *Receiver Operating Characteristic* (ROC) probability curve analysis. The drug–disease associations were ranked according to increasing p-values and a "real association" was assigned according to TTD information: 1 if the predicted drug-disease association is known, 0 otherwise. For a specified p-value threshold, the true positive rate (i.e., sensitivity) was calculated as the fraction of known associations that are correctly predicted, while the false positive rate (i.e., 1-specificity) was computed as the fraction of unknown associations that are predicted. The ROC probability curve was drawn based on these measures at different thresholds and the corresponding Area Under the Curve (AUC) was computed. Higher the AUC, better the algorithm is at distinguishing between two classes (i.e., known drug-disease associations *versus* unknown drug-disease associations).

### *Bi-Random walk-based algorithm*

Inspired by an already existing algorithm of drug repurposing called MBiRW [46], we exploited a Bi-Random walk-based (BiRW) approach to infer potential reuse for existing drugs. We started from the known drug–disease associations available on the TTD [19] and iteratively we added new associations integrating information retrieved from DrugBank [20], for what concerns targets of already approved drugs, and from Phenopedia [19], for what concerns disease genes. Denoting with $m$ the total number of the approved drugs and $n$ the total number of diseases, the known drug-disease associations were translated into a binary matrix $W_{rd}^{m \times n}$ by assigning 1 if the given drug (matrix row) is associated to a given disease (matrix column), and 0 otherwise. Next, a weighted adjacency matrix $W_{rr}^{m \times m}$ was computed to model drug similarity, where weights are the number of common targets between any pair of drugs. Further, a weighted adjacency matrix $W_{dd}^{n \times n}$ was computed to model disease similarity, where weights are the number of common disease genes between any pair of diseases (Figure 9a).

In analogy with [46], the matrices $W_{rr}$ and $W_{dd}$ were adjusted based on the known drug–disease associations. The underlying hypothesis is that drug (disease) pairs, whose similarity value is greater of what expected by chance, are more likely to share common diseases (drugs).

From the adjusted matrix $W_{rr}$, a weighted *drug similarity network* was built, in which nodes are the approved drugs and an edge occurs between two nodes if they share at least one target gene, with a weight given by the corresponding element of the matrix $W_{rr}$. Likewise, from the adjusted matrix $W_{dd}$, a weighted *disease similarity network* was built, in which nodes are the diseases, and an edge occurs between two diseases if they share at least one disease gene, with a weight given by the corresponding element of the matrix $W_{dd}$.

Next, new drug–disease associations were predicted through an iterative random walk process on drug and disease similarity networks, simultaneously. Denoting with the parameters $l$ and $r$ the walks in the drug network (left walks) and in the disease network (right walks), the bi-random walk can be described by the following equations:

$$A_t^l = \alpha W_{rr} A_{t-1} + (1-\alpha) A_0$$

$$A_t^r = \alpha A_{t-1} W_{dd} + (1-\alpha) A_0$$

where $A_0$ denotes the drug-disease association at $t = 0$ (matrix $W_{rd}$), while $A_t^l$ and $A_t^r$ represent the predicted drug–disease associations at iteration $t$ starting from left or right, respectively. The hypothesis behind is that an association between a drug R1 and a disease D1 can be added following these two avenues (Figure 9b):

1. a random walker starts from a random vertex R1 of the drug similarity network and in each step walks to one of the neighboring vertices (for instance R2, with a known or previously predicted association with D1) with a probability proportional to the weight of the edge traversed (i.e., the corresponding element of $W_{rr}$)

2. a random walker starts from a random vertex D1 of the disease similarity network and in each step walks to one of the neighboring vertices (for instance D2, with a known or previously predicted association with R1) with a probability proportional to the weight of the edge traversed (i.e., the corresponding element of $W_{dd}$)

In both cases, the existence of a known association between a drug R1 and a disease D1 has been considered with a probability $(1-\alpha)$. Then, in each step of the iteration process, the predicted drug–disease associations matrix $A_t$ is given by the mean between $A_t^l$ and $A_t^r$.

Finally, a bipartite drug-disease network was constructed consisting of two sets of nodes: one set corresponding to all disease, the other set corresponding to all approved drugs. An edge between a drug and a disease occurs if an association is already known or has been predicted among them.

## Tenfold cross-validation

The performance of BiRW-based algorithm in predicting new drug–disease associations was evaluated through a tenfold cross validation [46]. This is a technique to investigate the predictive power of an algorithm by partitioning the original sample into a training and test set. In particular, the tenfold cross-validation consists of randomly partitioning the original sample (in our case the drug-disease associations) into 10 equal size subsamples. Of the 10 subsamples, a single subsample is retained as test set and the remaining 9 subsamples are used as training set. The cross-validation process is then repeated for 10 iterations so that each of the 10 subsamples is used exactly once as the test set.

The results of each iteration were evaluated in terms of ROC probability curves. In particular, each time, the predicted drug-disease associations were ranked according to the similarity score estimated by the BiRW-based algorithm (i.e., the corresponding element of matrix $A_t$) and the $N$ top-ranked drug-disease associations were selected to be evaluated by the ROC curve analysis, where $N$ is the length of the test set. It's worth to stress that, in each cross-validation trial, we didn't use the information about the known drug–disease associations for the test set that were put to zero at first iteration of the bi-random walk process. Then, the ROC curve is constructed for different values of a specified threshold, where a true drug–disease association was considered as correctly predicted if the estimated similarity score of this association was higher than the specified threshold. The results of the ROC curve analysis obtained for each iteration were averaged to obtain a mean ROC curve and the corresponding Area Under the Curve (AUC) was calculated. Higher the AUC, better the algorithm is at distinguishing between two classes (i.e., known drug-disease associations *versus* unknown drug-disease associations).

### *Functional enrichment analysis*

Kyoto Encyclopedia of Genes and Genomes (KEGG) enrichment analysis aiming to evaluate functional pathways of SARS-CoV-associated disease genes was performed by using R statistical software and the package clusterProfiler [82]. P-values were adjusted with the false discovery rate (FDR) method and a threshold equal to 0.05 was set to identify functional annotations significantly enriched amongst the selected gene lists.

### *Gene set enrichment analysis*

In order to test whether the selected anti-SARS repurposable drugs can counteract the gene expression perturbations caused by the virus, (i.e., whether they down-regulate genes up-regulated by the virus or *vice versa*), we performed a gene set enrichment analysis (GSEA) as in [16]. We first collected gene expression datasets of hosts infected SARS-CoV available through the GEO public repository. In particular: transcriptome data of SARS-CoV-infected samples from patient's peripheral blood (GSE1739) and Calu-3 cells (GSE33267). To define differentially expressed genes, we selected adjusted p-values less than 0.05 and 0.01 for GSE1739 and GSE33267 dataset, respectively. These differentially expressed genes were used as SARS-CoV signatures, whereas the gene expression data of drug-treated human cell lines from the Connectivity Map (CMAP) database [39] were used as drug signatures. For each drug that was in both CMAP database and in our SARS-drug network, CMAP computed a score to evaluate the treatment effects of various drugs on genes that are hallmarks for SARS disease phenotype. Selected repurposing candidates with a score > 0 were considered to have potential treatment effect and the number of such SARS-CoV signature datasets was used as the final GSEA score that ranges from 0 to N, being N the total number of SARS-CoV signatures used.

## Acknowledgments

First of all, we would like to thank the health care workers that in this dramatic moment are on the front lines giving everything to fight this novel coronavirus pandemic. We would like to sincerely thank Dr. Gianpiero D'Offizi for his precious advice on drugs and drugs combination to include in our analysis and for his important help in the interpretation of our results.

This work was financially supported by PRIN 2017 - Settore ERC LS2 - Codice Progetto 20178L3P38 and by Sapienza University of Rome grant entitled "Network medicine-based machine learning and graph theory algorithms for precision oncology" - n. RM1181642AFA34C2.

## Author contributions

PP, LF conceived and designed the research. PP, GF, FC developed the algorithm and contributed to computational data analysis. All authors contributed to interpretation of data, writing, and approval of the final manuscript.

## Conflict of interest

The authors declare that they have no competing interests.

## Additional files

**SupplementaryTable1.xlsx:** the table is composed of two separate sheets. The first sheet reports the diseases analyzed in our study with the corresponding number of disease-causing genes obtained from Phenopedia database. The second sheet reports the FDA-approved drugs obtained from DrugBank and processed in our analysis with the corresponding number of target proteins.

**SupplementaryTable2.xlsx:** the table is composed of three separate sheets. The first sheet reports the bipartite SARS-drug subnetwork released by SAveRUNNER; the second sheet reports the bipartite COVID-19-drug subnetwork released by SAveRUNNER; the third sheet reports the lists of specific and common candidate repurposable drugs for COVID-19 and SARS.

# Figures

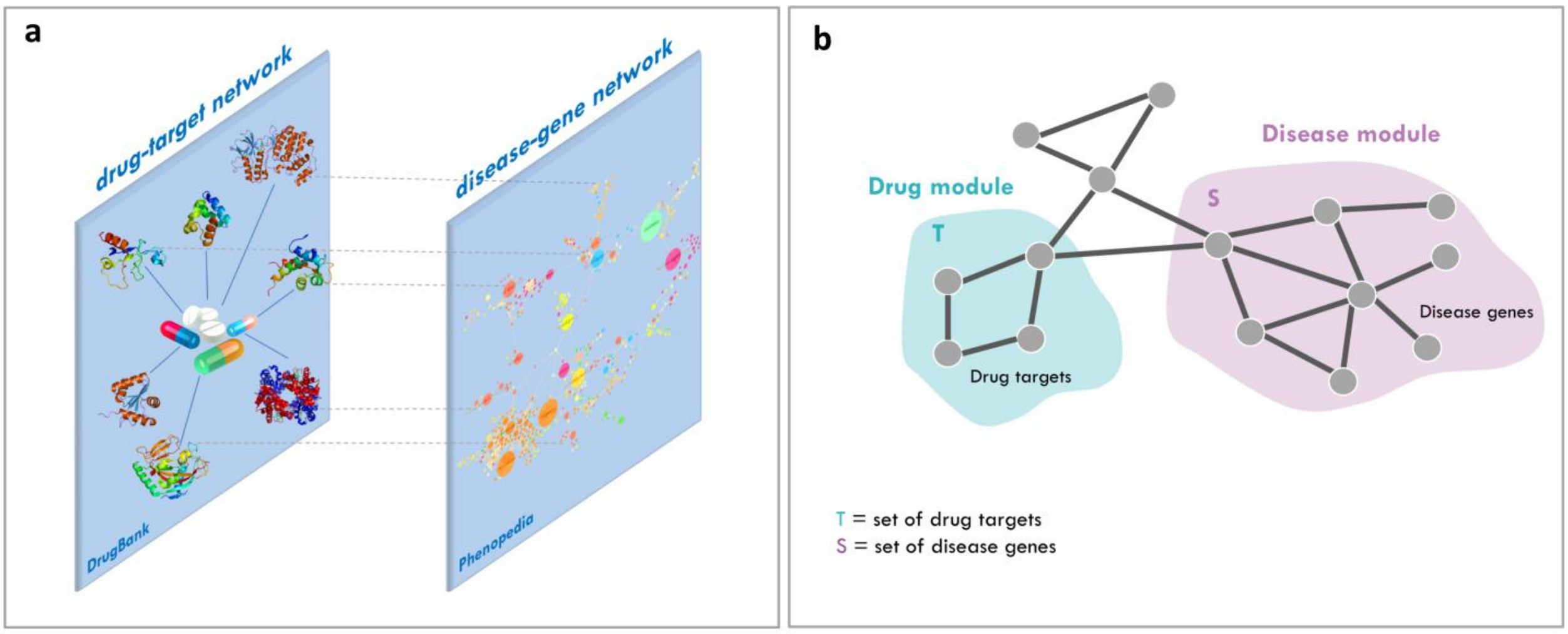

**Figure 1. Schematic representation of SAveRUNNER inputs and working hypothesis.** (a) *Inputs*. SAveRUNNER takes as input the list of drug targets downloaded from DrugBank database and the list of disease genes downloaded from Phenopedia database. These lists can be represented as networks: (i) a drug–target network, within which nodes are drugs and target proteins, linked if the protein is a known target of the drug; and (ii) a disease-gene network, within which nodes are diseases and genes, linked if the gene has been associated to the disease (b) *Working hypothesis*. Potential candidate repurposable drugs for a given disease should have target proteins (drug module T) within or in the immediate vicinity of the disease module S.

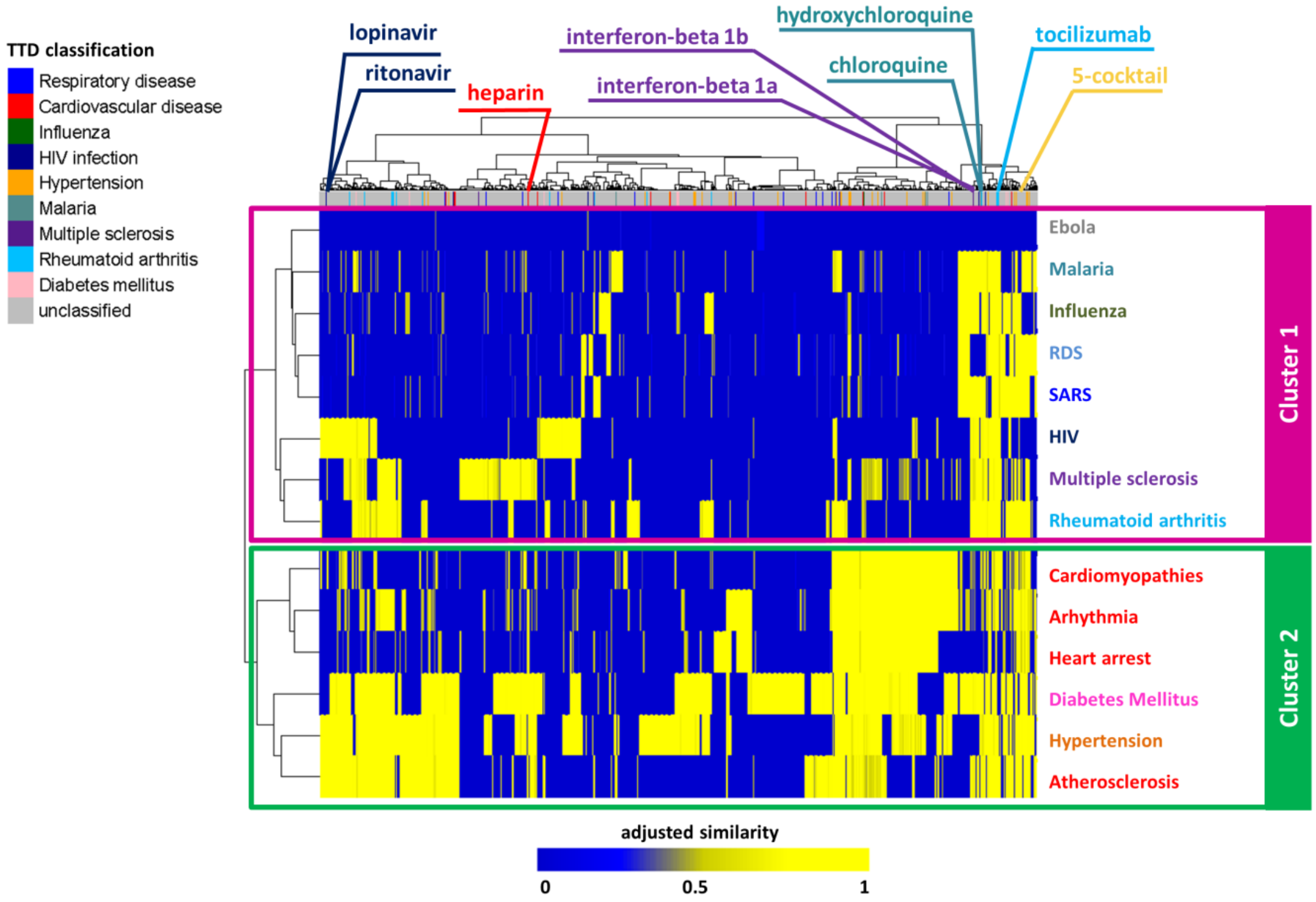

**Figure 2. Dendrogram and heatmap of the drug-disease network.** The drug-disease network similarity values are clustered according to rows (diseases) and columns (drugs) by a complete linkage hierarchical clustering algorithm and by using the Euclidean distance as distance metric. Heatmap color key denotes the adjusted and normalized network similarity between drug targets and disease genes in the human interactome, increasing from blue (less similar) to yellow (more similar). Drugs are colored according to the Therapeutic Target Database (TTD) indications listed in legend. Unclassified tag was assigned to those drugs for which a known therapeutic indication was not available in TTD or their indication does not fall in our analyzed disorders.

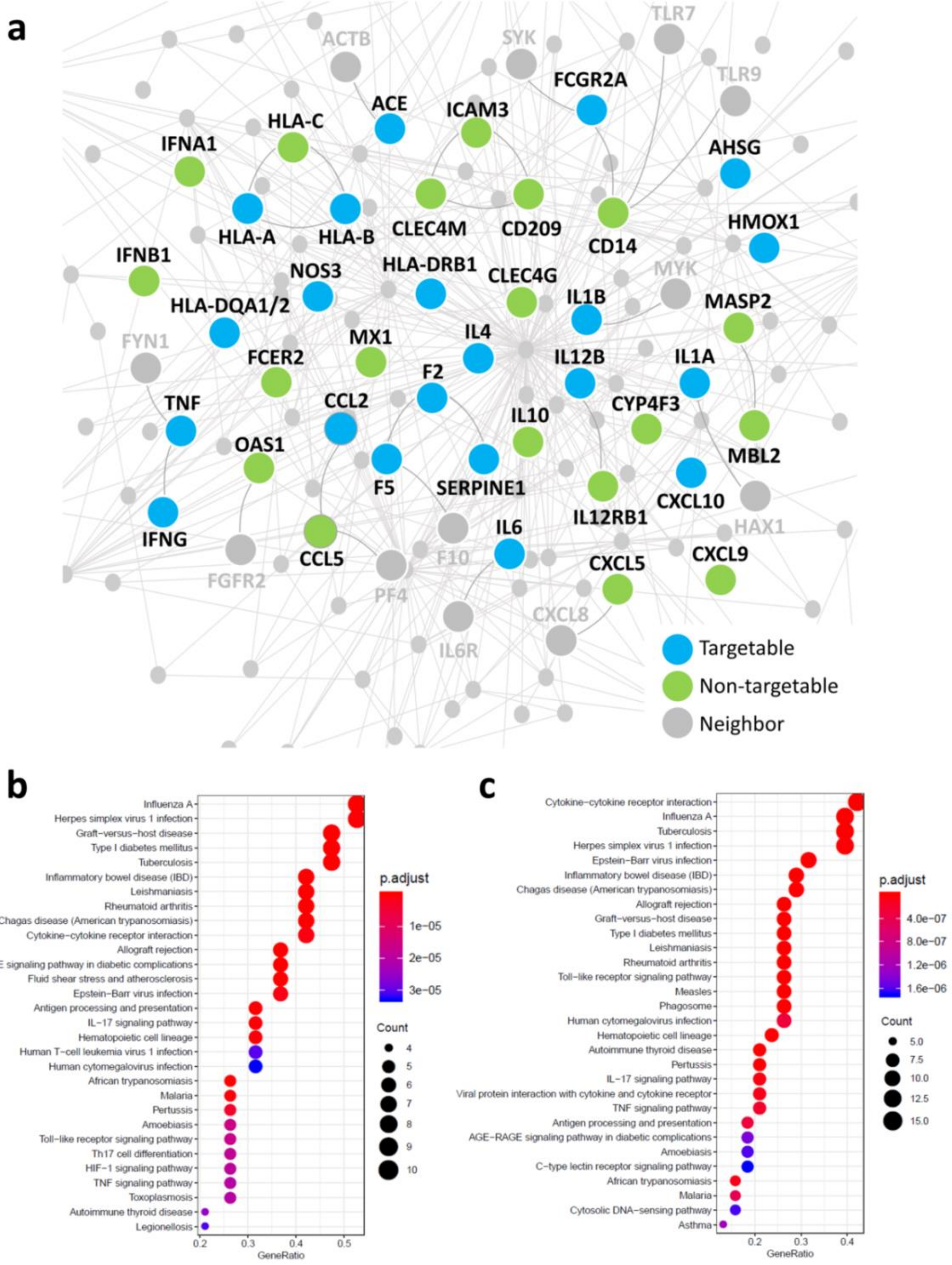

**Figure 3. SARS-CoV-host interactome.** (a) *The SARS-CoV-associated disease genes subnetwork in the human interactome.* Light blue nodes represent SARS-CoV-associated proteins that can be directly targeted by at least one FDA-approved drugs (targetable); green nodes represent SARS-CoV-associated proteins that do not have any known ligands and then cannot be directly targeted by drugs (non-targetable); grey nodes represent interaction partners of SARS-CoV-associated proteins in the human interactome (neighbor). (b-c) *KEGG human pathway enrichment analysis for SARS-CoV-associated disease genes*. The dot plots of the top 30 enriched KEGG pathways (p-value ≤ 0.05) obtained for the 21 targetable SARS-CoV-associated disease genes (b) and for the total 41 SARS-CoV-associated disease genes (c). The y-axis reports the annotation categories (KEGG pathways) and the x-axis reports the gene ratio (i.e., the number of genes found enriched in each category over the number of total genes associated to that category). The color of the dots represents the adjusted p-values (FDR), whereas the size of the dots represents the number of genes found enriched in each category.

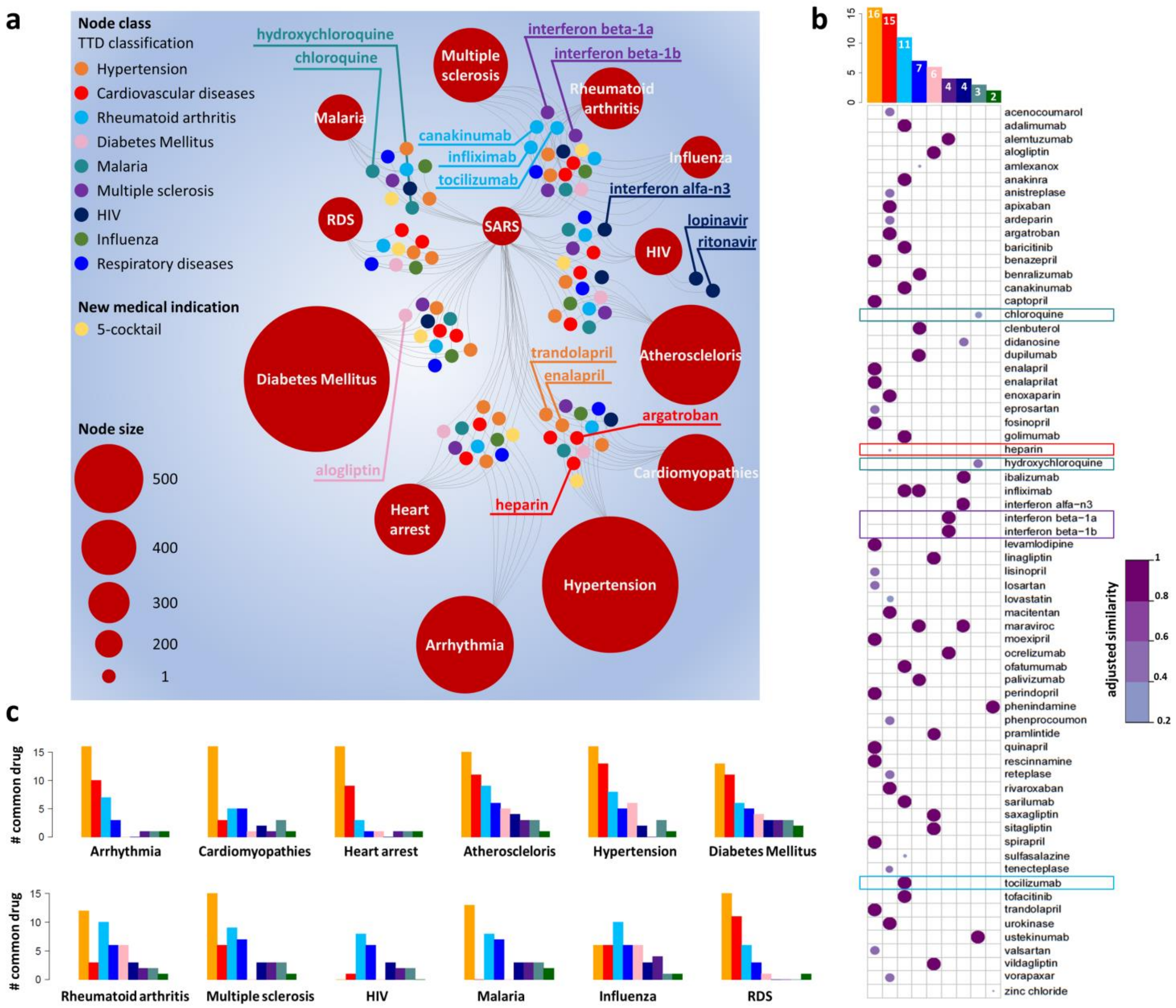

**Figure 4. The predicted SARS drug-disease network.** (a) *Schematic representation of the SARS predicted drug-disease network***.** This sketch shows the high-confidence predicted drug-disease associations connecting SARS and other analyzed diseases (red circles) with the 66 FDA-approved non-SARS drugs or the new proposed medical indication (i.e., 5-cocktail). Drugs are colored according to TTD classification reported in the legend, or according to the new proposed medical indication (i.e., 5-cocktail). The node size scales indicate the degree (connectivity) of nodes in the predicted drug-disease network. Labeled drug nodes represent either drugs more significantly associated to SARS or drugs being currently explored as COVID-19 treatment. (b) *Similarity plot.* Network-predicted repurposable drugs for SARS (along rows) with their TTD classification (along columns). In the plot, circles are scaled and colored according to the adjusted similarity measure, increasing from light purple (low similarity) to dark purple (high similarity). The barplot placed on the top reports the total number of candidate repurposable drugs for SARS grouped and colored according to TTD classification. (c) *Common drugs between SARS and other diseases.* For each analyzed disease, the barplot reports the total number of drugs shared with SARS grouped and colored according to TTD classification.

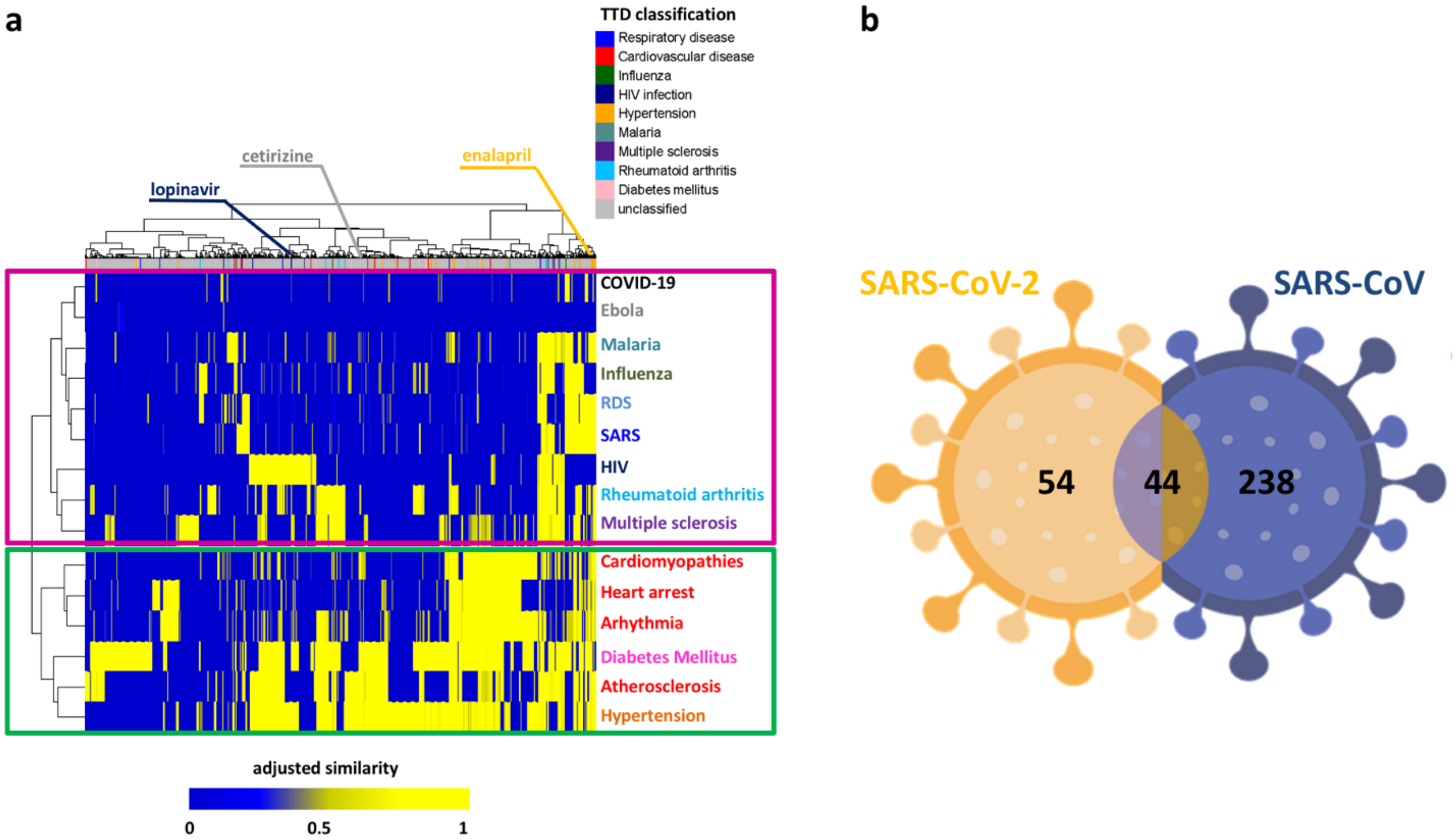

**Figure 5. Candidate repurposable drugs for SARS-CoV-2.** (a) *Heatmap and dendrogram of SARS-CoV-2 drug-disease network***.** The statistically significant (p-value ≤ 0.05) network adjusted similarity values are clustered according to rows (diseases) and columns (drugs) by a complete linkage hierarchical clustering algorithm and by using the Euclidean distance as distance metric. Heatmap color key denotes the adjusted similarity between drug targets and disease genes in the human interactome, increasing from blue (less similar) to yellow (more similar). Drugs are colored according to the Therapeutic Target Database (TTD) indications listed in legend. Unclassified tag was assigned to those drugs for which a known therapeutic indication was not available in TTD or their indication does not fall in our analyzed disorders. (b) *SARS-CoV-2 versus SARS-CoV*. Venn diagram detailing the number of common and specific candidate repurposable drugs predicted by SAveRUNNER for SARS-CoV-2 and SARS-CoV infections.

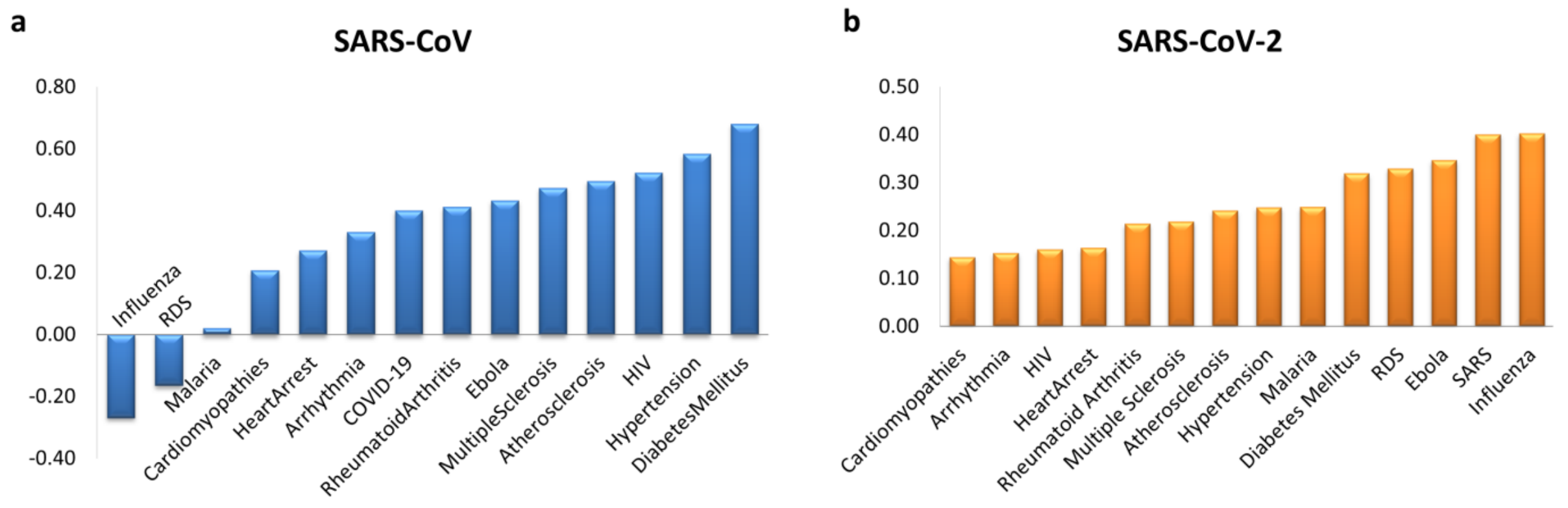

**Figure 6. Disease comorbidity.** The bar plots show the values of the non-Euclidean separation distance computed for the SARS-CoV (a) and SARS-CoV-2 (b) disease module with respect to all the other disease modules analyzed in this study.

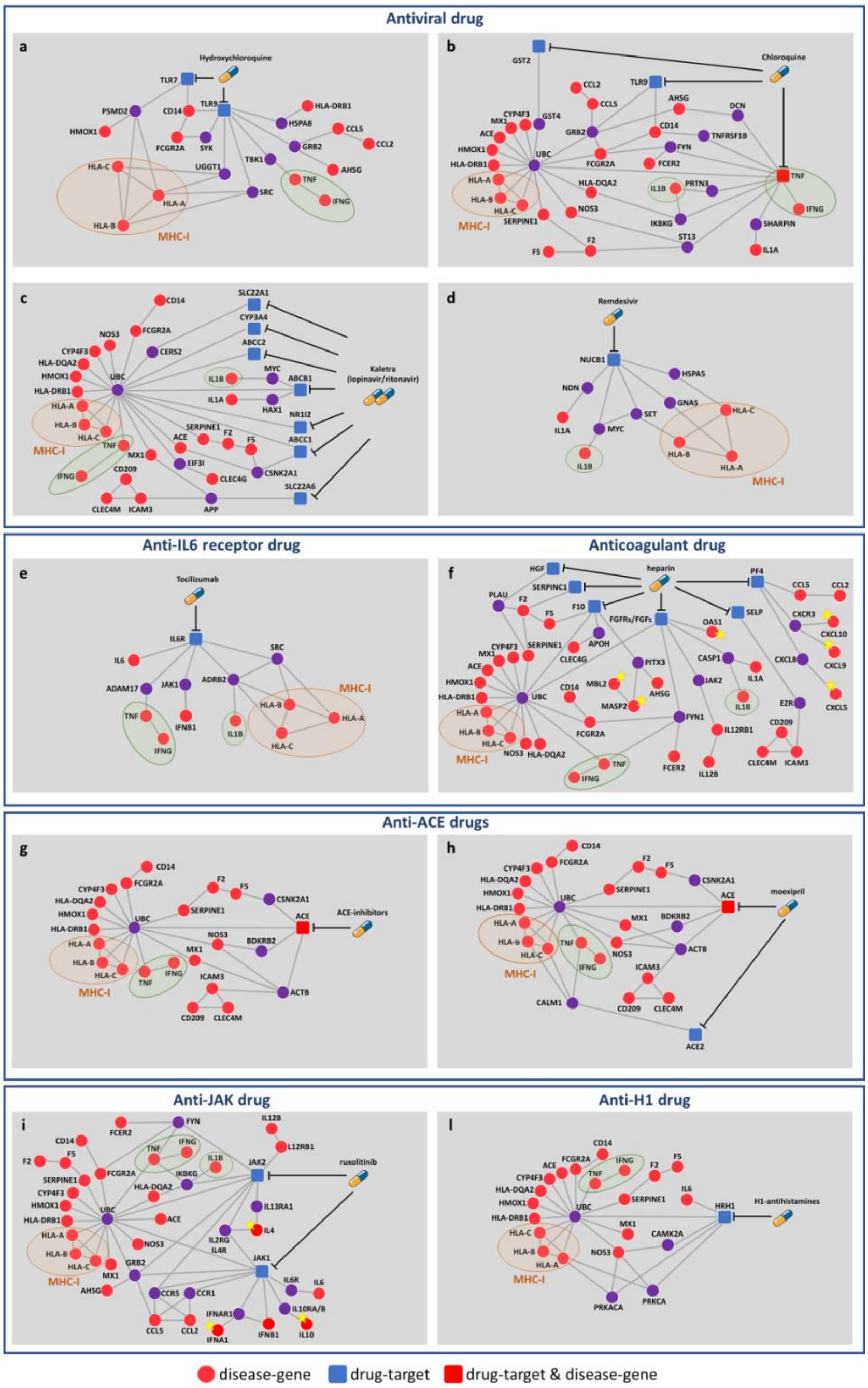

**Figure 7. Mechanism-of-action of anti-SARS-CoV repurposable drugs.** The subnetworks show the inferred mechanism-of-action for: antiviral drugs (a-d), *tocilizumab* (e), *heparin* (f), ACE-inhibitors (g-h), *ruxolitinib* (i), and H1-antistamines (l). Each subnetwork was designed to point out the shortest paths from drug targets and SARS disease genes in the human interactome. In each subnetwork, disease genes specifically targeted by each drug are marked with a yellow star; major histocompatibility complex of class I (MCH-I) and pro-inflammatory cytokines shared by all drug networks are marked with an orange and a green circle, respectively. Legend: red circles refer to SARS disease genes, blue squares refer to drug targets, red squares refer to SARS disease genes that are also drug targets, violet circles refer to the first nearest neighbors (that are not disease genes) of the drug targets in the human interactome. Anti-ACE drugs of panel (g) refer to *enalapril*, *trandolapril*, *fosinopril*, *benazepril*, *cilazapril*, *zofenopril*, *spirapril*, *rescinnamine*, and *quinapril*. H1-anistamines of panel (l) refer to *fexofenadine, levocetirizine, desloratadine, clemastine, cetirizine*.

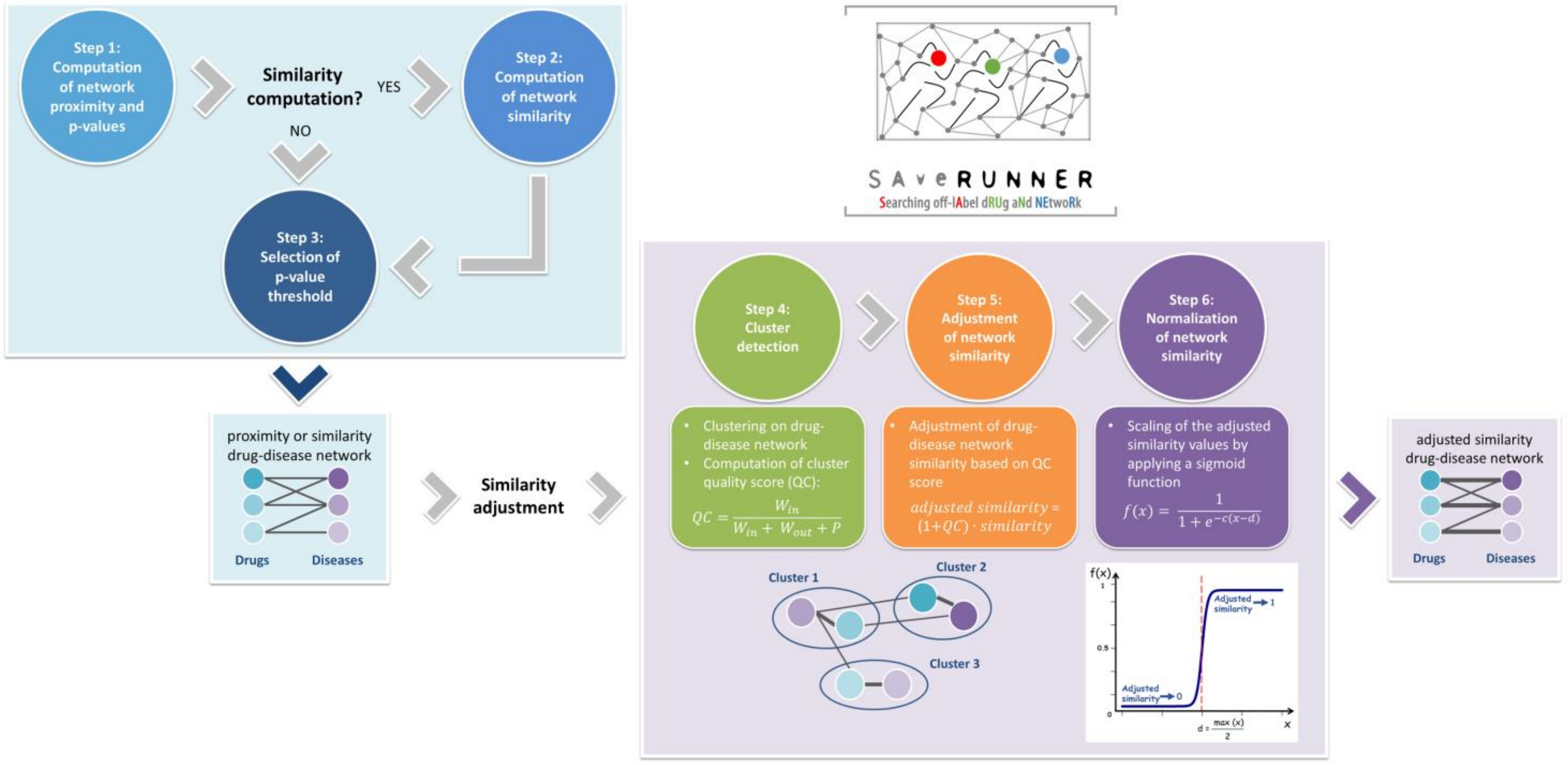

**Figure 8. SAveRUNNER algorithm.** SAveRUNNER encompasses six steps: (1-3) compute a weighted bipartite drug-disease network, where nodes are both drugs and diseases, edges are the statistically significant drug-disease associations (p-value ≤ selected threshold), and weights are either the proximity or similarity measure; (4-6) compute the normalized adjusted similarity measure to correct the weights of the drug-disease network and to prioritize the predicted drug-disease associations. Legend: QC is the quality cluster score; $W_{in}$ is the total weight of edges within each cluster; $W_{out}$ is the total weight of edges connecting each cluster to the rest of network; *P* is the node density within each cluster; *c* and *d* parameters are the sigmoid steepness and midpoint, respectively.

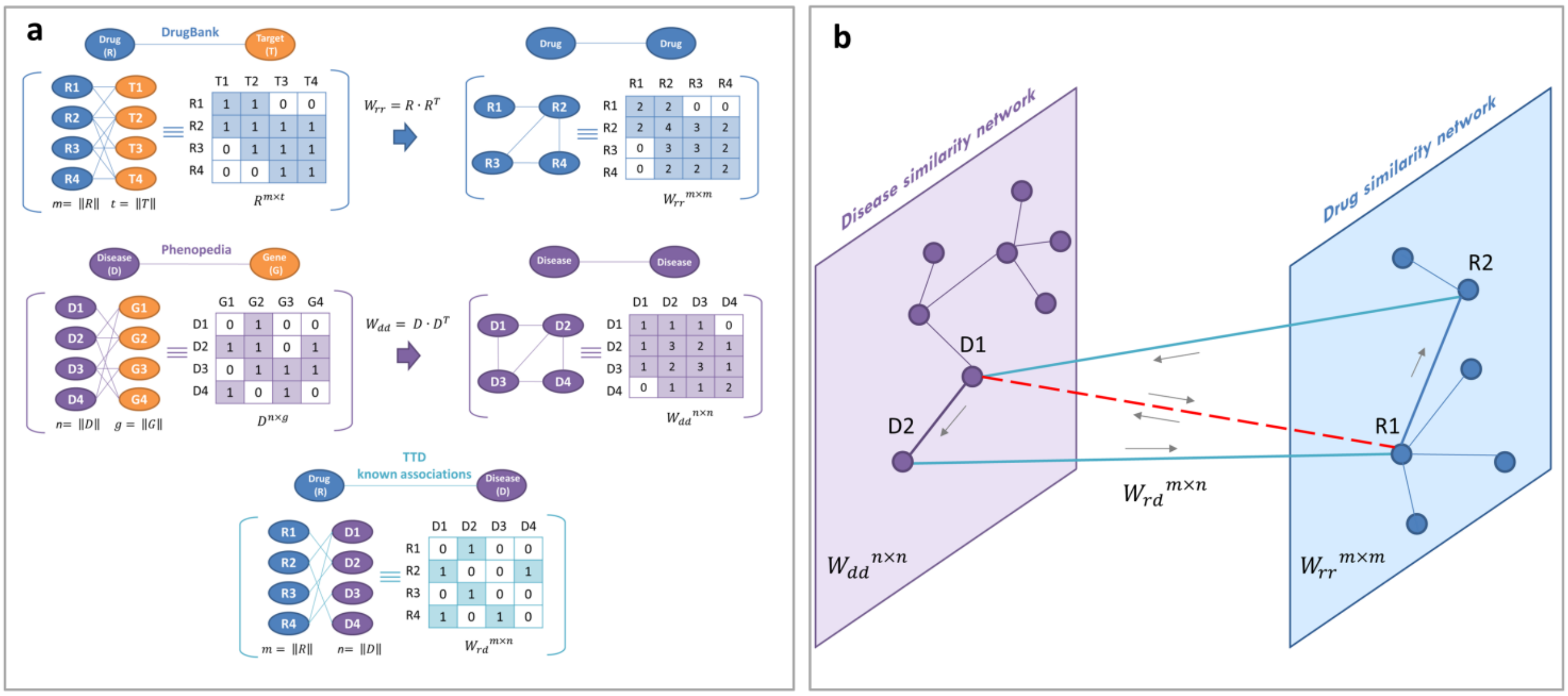

**Figure 9. BiRW-based algorithm**. Schematic representation of the construction of input matrices (a) and predicted drug-disease associations network (b).